\documentstyle[11pt,aaspp4]{article}

\slugcomment{In press in The Astronomical Journal}

\def\mv{M$_V^{HB}~$}
\def\fe{~[Fe/H]~}
\def\vs{{\it vs.~}}

\def\etal{{\it et~al.~}}
\def\hst{{\it ~HST~}}

\def\dmo{~(m-M)$_o$~}
\def\dv{~$\Delta V^{TO}_{HB}$~}
\def\vhb{~V$_{HB}$~}
\def\drich{$\Delta V_{HB}^{rich}$}

\lefthead{Fusi Pecci et al.}
\righthead{The \mv \vs \fe Calibration: I.}

\begin{document}

\title{ The \mv \vs \fe Calibration: I.\\
        HST Color-Magnitude Diagrams of\\
        8 Globular Clusters in M 31}

\author{F. Fusi Pecci\altaffilmark{1}, R. Buonanno\altaffilmark{2},
C. Cacciari\altaffilmark{1},}
\author{C.E. Corsi\altaffilmark{2}, S.G. Djorgovski\altaffilmark{3},
L. Federici\altaffilmark{1},}
\author{F. R. Ferraro\altaffilmark{1}, G. Parmeggiani\altaffilmark{1},
R.M. Rich\altaffilmark{4}}

\altaffiltext{1}{Osservatorio Astronomico di Bologna, Via Zamboni 33,
40126 Bologna, Italy}
\altaffiltext{2}{Osservatorio Astronomico di Monte Porzio, Roma, Italy}
\altaffiltext{3}{Division of Physics, Mathematics, and Astronomy, California
Institute of Technology, Pasadena, CA 91125, USA}
\altaffiltext{4}{Department of Astronomy, Columbia University, New York, 
NY 10027, USA}

\begin{abstract}
Color Magnitude Diagrams (CMDs) of individual stars in 8 Globular
Clusters (GCs) in M31 down to about 1 mag fainter (V$\sim 26.5$)
than the Horizontal Branch (HB) have been obtained with the 
{\it Hubble Space Telescope -HST}.

In particular, we observed G280 and G351 with the FOC ($f/96+F430W$ and 
$f/96+F480LP$) while the WFPC2 (F555W, F814W) frames for G1, G58, G105, G108, 
G219+Bo468  were retrieved from the \hst archive.

The cluster metallicities -\fe- range from -1.8 to -0.4. Coupled with
sufficiently accurate (to $\sim\pm0.1$ mag) measures of the mean brightness of 
the HB --\vhb--, appropriate estimates of reddening for each
cluster, and the adoption of a distance modulus to M31 of \dmo = 24.43,
this has allowed us to yield a direct calibration for the mean absolute
magnitude of the HB at the instability strip --\mv-- with varying \fe:
\begin{displaymath}
M_V^{HB} = (0.13 \pm 0.07)\fe + (0.95 \pm 0.09)
\end{displaymath}
where the associated errors result from the adopted global errors in the 
measure and best fitting procedures.

The slope of the derived relation is fully consistent with that predicted
by the standard and canonical models ($\sim0.15$) and obtained by various
ground-based observations, while it is only marginally compatible 
with higher values ($\sim0.30$), also obtained in the past.

The zero-point, which is crucial to {\it absolute} age determinations, 
depends on the adopted distance to M31 and is moreover affected 
by an additional error due to the residual uncertainties in the
\hst photometric zero-points ($\sim0.05$ mag, at least).

If confirmed, such a calibration of the \mv \vs \fe relationship
would imply {\it old} absolute ages ($> 16$Gyr) for the oldest Galactic 
globulars and fairly small age spread among those having a
constant magnitude difference between the Main-Sequence Turnoff and
the HB.
\end{abstract}


\section{Introduction}
Horizontal Branch (HB) stars are fundamental standard candles for Population
II systems, and consequently are important tools for deriving ages 
of globular clusters (GCs) from the Main - Sequence Turnoff (TO) luminosity.

While other techniques exist to calibrate the TO luminosity, they require
high photometric precision at faint magnitudes, so the use of the
magnitude difference between TO and HB --\dv-- 
remains, in our opinion, one of the best approaches
to age determination for most of the
Galactic GCs. However the conflict between the large ages estimated for 
globular clusters and the smaller values derived from the 
expansion models is a source of considerable concern for standard 
cosmological studies. 

The luminosity level of the HB --\mv--, namely its absolute ``vertical''
location in the HR diagram, is expected (based on stellar evolution
theory) to depend on metallicity --\fe.  Despite considerable effort over the
last decade, the slope of this relation is still not known very accurately,
ranging from $\sim 0.15$ to $\sim 0.37$ (see \cite{san93} for
a recent review).  
The zero point is also uncertain at the 0.25 mag level,
and these uncertainties propagate into an error of at least $\pm 3$ Gyr
in the determined absolute ages of Galactic GCs. Therefore 
the current uncertainties are too large and make it impossible to
assess if the GGCs have an age-metallicity relation.  

The age calibration of the observable parameter \dv 
(\cite{buo89,scacc90,wal92,bolte96,cha96,buo96}, Catelan \& de Freitas 
Pacheco 1996)
critically depends on the \mv \vs \fe relation. Since \dv has been measured 
for several GGCs spanning from the outer halo to the Galactic Bulge, 
a firm knowledge of this relation will have 
a widespread impact on the age question in general, and on the models of 
galaxy formation in particular. 

By extending to the nearest giant spiral galaxy M31 the same type of studies 
of the GC family that have become classical for our own Galaxy, 
it is possible to solve the problem of the precise measurement of the 
absolute luminosity of HB stars with varying metallicity, being the clusters 
in practice at the same distance to us.

The M31 globular cluster system offers a unique opportunity for stellar
population studies (see for references \cite {bro93,fp93,hu93}).
The cluster population is extremely rich, and the proximity of M31 allows the 
horizontal and giant branch populations to be resolved with \hst.
Most of the bright M31 clusters are not
seen projected against a disk population (as is the case for M33).
Almost all the clusters appear to be as old as those of the Milky
Way halo, with a wide range of metallicities (which represents a significant 
difference from the case of the Magellanic Clouds).
In M31 we have the further advantage that the distance dispersion  is at 
most $\Delta$\dmo$\sim 0.15$mag (corresponding to a maximum intrinsic 
depth of the M31 GC system of about 100 Kpc) 
and the reddening distribution is rather uniform, allowing 
comparative studies that are impossible to undertake for Milky Way clusters.

Beginning with \cite {hea88} and continuing
to present samples (\cite{cou95}) even the
best ground-based efforts using CFHT have failed to reach the
HB and resulted in photometry of a handful of the brightest and most external 
stars in each cluster.
Hence, \hst is the only instrument presently capable of carrying on 
this project.

This program was proposed as a GTO Program (no. 1283; subsequently
withdrawn) and approved as GO 5420 in Cycle 4. 
Data from FOC/96 obtained during Cycle 1 (GO 2583) 
has resulted in publication of excellent surface brightness profiles 
for 13 clusters (\cite{fp94})
including the discovery of the 
first post-core collapse M31
globular cluster, which could not have been discovered from the ground
(\cite{ben93}).

Taking full advantage of the post-refurbishment characteristics, 
at the end of Cycle 5 a total of 12 M31 clusters will have been
observed with WFPC2 or COSTAR+FOC.  We present here the CMDs obtained
from the reduction and analysis of 8 clusters. Two of them have been
observed by us with the FOC, and for the other six we retrieved the available
archive WFPC2 data. Table 1 lists the program clusters as identified 
in the various catalogs (G = \cite{sar77}, Bo = \cite{bat93},
 V = \cite{vet62}) and a few basic data.
\placetable{tbl-1}

\section {Observations and reductions}

The full description of the detailed photometric treatments and the data 
is postponed to a forthcoming paper where also the surface 
brightness profiles and 
the structural cluster parameters will be measured and discussed.

\subsection {FOC-data}

The completely refurbished and properly aligned and focused
\hst+COSTAR system with the high resolution Faint Object Camera (FOC) 
offers in principle a powerful tool to study the GCs in the distant
galaxies of the Local Group. In particular, the M 31 GCs are the ideal 
targets since at this distance ($\sim 770$ Kpc) the typical dimension of a 
GC ($\sim 25$pc) corresponds to only $7''$, and this perfectly fits into 
the FOC field of view. 
In addition, the FOC is the \hst camera with the highest space resolution, 
which is desirable for the photometry of individual stars 
and cannot be matched by the space resolution of ground based observations
even in excellent seeing conditions.

The data presented here were obtained on 18-19 January 1995
and 23 January 1995, for the M31 GCs G351 and G280, respectively,
using the $512\times512$ pixel$^2$ imaging mode of the f/96 camera of
the COSTAR-corrected FOC. In this configuration the pixel size is
$0".0144\times 0".0144$ and the field of view is $\sim 7"\times 7"$
(\cite{jed94}). Five exposures ($4\times2300+1\times1600$~sec) 
were obtained using the F480LP filter and two ($1500+2300$~sec)  
using the F430W on each cluster.

The raw images were flattened and geometrically corrected 
using the standard STScI pipeline calibration procedure in order to remove
the optical distorsions. 

All images in each filter have been registered and co-added in order
to increase the $S/N$ of the faint stars. The final images ready for
reduction correspond to an effective exposure time of 10,783 sec and 3793.2
sec in the F480LP and F430W filter, respectively.
Figure 1 presents the result of this procedure for G351 (F480LP). 
\placefigure{fig1}

The standard searching procedure available in ROMAFOT (see Ferraro \etal 1991)
was then applied to each frame, excluding a small central area (with $r<130$ 
and $r<150$ pixels corresponding to $1.9$ and $2.2$ arcsec for G351 and G280, 
respectively) in order to avoid regions which are too severely crowded
even with \hst. More than 800 objects have been identified in each cluster
in the (deepest) F480LP frames, and these identifications were subsequently
used as input for the fitting procedure also in the F430W frames.
An interactive check of all the objects was carried out in the F430W
frames over a wide annulus to increase the degree of completeness of 
at least the brightest sample.

The photometry of individual stars on the co-added frames was performed
using ROMAFOT (\cite{buo83}), purposely adapted to handle \hst data.
In particular,  the \hst point spread function (PSF) has been modelled
by a \cite{mof69} function in the central part of the profile 
plus a numerical experimental map of the residuals in the wings. 
The optimal PSF has been determined from the analysis  of the brightest 
uncrowded stars independently in both co-added frames. 

The instrumental magnitudes resulting from the PSF fitting procedure have been 
transformed to the {\it HST} photometric system by using a sample of selected 
isolated stars whose magnitudes have been computed using both the fitting 
procedure and the
classic aperture photometry. For these reference stars \hst magnitudes 
have been computed using the formula (HST Handbook Data p.116, and Nota 1995,
private communication):
\begin{displaymath}
Mag_{HST} = -21.10 - 2.5Log((c\times U)/(\epsilon\times t))
\end{displaymath} 
where $c$ represents the stellar counts, $U$ is the inverse sensitivity 
function, $t$ is the exposure time in sec, and $\epsilon$ is the fraction 
of the encircled energy within the area considered.

On the basis of the values of the encircled energy within 2 pixels (radius)
reported in Table 8 of the FOC Handbook (Version 5.0, May 1994), we adopted
$\epsilon_{F430W}=0.46$ and $\epsilon_{F480LP}=0.42$.
For the inverse sensitivity function we have used
$U_{F430W}=0.28129\times 10^{-17}$ and
$U_{F480LP}=0.57939\times 10^{-17}$ erg cm$^{-2}$ A$^{-1}$. 

To construct an empirical transformation from the \hst magnitude system 
to the Johnson standard system, we have chosen a set of (V,B-V) points 
appropriate to define the fiducial Red Giant Branch 
(RGB) and HB for a wide sample of Galactic GCs. To each (V,B-V) pair 
we have then associated a simulated stellar spectrum of appropriate 
luminosity and spectral type generated from the \cite{kur79} stellar
atmosphere models.
Finally, by running FOCSIM (W.J.Hack, The FOCSIM Beginners Manual,
Version 2.0,May 1993)
we have obtained the following transformations:
\begin{displaymath}
B = m^{0}_{F430W} - 0.43\times(m_{F430W} - m_{F480LP})^{0} + 0.51
\end{displaymath}
\begin{displaymath}
V = m^{0}_{F480LP} - 0.20\times(m_{F430W}-m_{F480LP})^{0} + 0.04
\end{displaymath}
where $(m_{F430W}-m_{F480LP})^0$ is the dereddened color in the \hst
system, and $m^0_{F430W}$ and $m^0_{F480LP}$ are the dereddened \hst 
magnitudes computed by adopting $A_{F480LP}=3.46\times E(B-V)$ and    
$A_{F430W}=4.47\times E(B-V)$ (\cite{sav77}).
The calibration curves in each color are plotted in Figure 2a,b. 
\placefigure{fig2}
This procedure assumes that the reddening E(B-V) is known. See Sect. 3.3
for a description of our reddening assumptions. 

\subsection {WFPC2-data}

The images used in the present study have been retrieved from
the \hst archive. They consist of the data obtained in February 1994
(prior to the cooldown of the WFPC2) on G58, G105, G108 and G219 with 
the GTO program 5112 (P.I.: J. Westphal) and those obtained on G1 as a
part of Cycle 4 program GO 5464 (P.I.: M. Rich) in July 1994
(after the cooldown). In addition, we have reduced and analysed also 
the small globular cluster Bo468 which fell in the WF3 frame of the 
exposures taken primarily for G219. 
The results obtained from this material
have been published by \cite{ajh96} and \cite{ric95},
respectively, except for Bo468 which has never been studied before. 

Although the CMDs presented by these two groups are of excellent quality, 
we believed that a new independent analysis of the
whole data set was worth the effort, in order to combine 
all the available information on the \hst CMDs of M31 GCs
in the most homogeneous way as possible. Of course,
we have the problem of comparing data from two different instruments, 
with the additional complication that some WFPC2 data were taken before 
and some after the camera was cooled down from -77 to -88$^o$C. 
Nevertheless, the use of the same reduction procedure will at least
minimize the differences in the photometric treatments, and we believe we 
have taken as much care as possible to account for the intrinsic and 
systematic differences between the instruments. 

In particular, we have reduced all the PC frames taken from GTO 5112 
(2$\times$1000s with F555W (V) + 2$\times$1000s with F814W (I)) 
and the related WF3 frames which contained Bo468, 
and the PC frames taken with the same filters from GO 5464
(1600s with F555W (V) + 1200s with F814W (I), in total).
Concerning the pre-processing and calibrating procedures, since we
have simply retrieved the data from the \hst Archive,
we have strictly adopted the pipeline applied at the STScI.

The analysis of the available frames has been carried out following
two main methods, {\it i.e.} by cosmic-ray cleaning and co-adding 
the frames to yield a master image in each color on which the photometric 
measurements were performed, or by independently reducing and measuring 
all the frames and subsequently matching and averaging the individual
sets to yield a final cross-checked sample. 
To our experience, the first method yields slightly deeper samples
(by about half a magnitude), whilst the second yields a better photometric
accuracy for the objects down to about one magnitude above the frame limit.

As reported above for the FOC data, the study of the PSF, the search 
for the objects, and the measure of the instrumental magnitudes
have been carried out by using ROMAFOT (\cite{buo83}),
excluding a small central area (with $r<30$ pixels corresponding to 
$\sim 1.5$ arcsec for 
G58, $r<45$ pixels corresponding to $\sim 2$ arcsec for G105, G108, G219, and
$r<15$ pixels corresponding to $\sim 1.5$ arcsec for Bo468)
in order to avoid regions where the crowding is too severe even with \hst. 
About 1987, 1216, 555, 1228 (+ 1466 in the field), 726 and 335 (+ 1580
in the field) stellar objects have been identified 
in G1, G58, G105, G108, G219 and Bo468 respectively, in the (deepest) 
F814W frames. These identifications were subsequently used as input for 
the fitting procedure also in the F555W frames.

The instrumental magnitudes resulting from the fitting procedure have been 
transformed to the \hst ~photometric system by using a sample of selected 
isolated stars whose magnitudes have also been computed using the
classic aperture photometry after correcting for CTE effects (\cite{hol95}).
For these reference stars \hst magnitudes 
have been computed using the formula (\cite{hol95}):
\begin{displaymath}
V = F555W - 0.052(V-I)_{0} + 0.027(V-I)_{0}^{2} + 21.725 + 2.5 log_{10}(1.987)
\end{displaymath}
\begin{displaymath}
I = F814W - 0.062(V-I)_{0} + 0.025(V-I)_{0}^{2} + 20.839 + 2.5 log_{10}(1.987) 
+ c
\end{displaymath}
where $c$ has been set equal to +0.05 for the observations taken from 
GTO 5112 (with the detector operating at -77$^o$C) and equal to 0.00
for G1 (from GO 5464, at -88$^o$C) as done by \cite{ajh96}
and \cite{ric95}, respectively.

\subsection {Photometric errors and comparison with previous 
photometries}

Both \cite{ajh96} and \cite{ric95} have widely discussed
the problem of estimating the global ({\it i.e.} internal + systematic)
errors in the \hst photometry of the M31 GCs, also using simulations and 
artificial star experiments.
A detailed discussion of our photometric errors and a complete
comparison with the previous photometries will be presented
in the forthcoming full paper. We report here schematically the
main conclusions and figures.

The {\it internal} errors are mostly due to noise from the sky, 
imperfections of the PSF fitting, and blending effects. The method
adopted with ROMAFOT to describe the PSF is particularly effective
in reducing the size of the internal errors essentially because of
the combination of a "known" Moffat-function and a map of the local
residual plus a multicomponent fitting procedure. A rough quantitative
estimate of the internal errors can be obtained by computing
the r.m.s. scatter of the individual measures for the same stars
on different frames. Adopting the formula taken from \cite{fer91} 
one has 
$\sigma_V\sim0.02$~mag and $\sigma_I\sim0.02$ mag in the
interval $22<$V$<24$ and
$\sigma_V\sim0.05$~mag and $\sigma_I\sim0.06$ mag in the
interval $24<$V$<27$.

Another straightforward way of estimating our internal errors
can be the direct comparison of the observed widths of the various
branches in the final CMDs for the clusters in common with
previous studies based on the same original \hst data.
Figure 3a,b compares the CMDs of G 105 as obtained here and
as reported in Figure 9 of \cite{ajh96}. 
\placefigure{fig3}
As one can see,
both the location and the spread of the points around the main
ridge lines are comparable, and even the substructures along
the branches are very similar. This ensures that the photometric
quality of the reductions are comparable, and we conclude that
our {\it internal} photometric errors are not larger than 
those quoted by \cite{ajh96}, which they claimed to
be fully consistent with those expected from their simulations.

Concerning the {\it systematic} errors, there are many reasons of
uncertainty and most of them can hardly be quantified safely.
In particular, there are problems related to charge transfer efficiency 
(CTE) with WFPC2. Then, there
are uncertanties in the conversion to Standard System for
both cameras. Finally, there is additional concern about the
exact zero-points of the HST photometric systems. 

We do not have any special information and new data on these items and 
refer to the thorough discussions reported for the two cameras 
by Holtzman \etal (1995), \cite{ajh96}, \cite{coo95,ric95}. 

In conclusion, we believe that while the {\it internal errors}
are as close as possible to the minimum limit achievable, 
the residual {\it systematic}
errors, especially in the zero-points, can still be quite large
(up to 0.05 mag, \cite{ajh96}). This may have a strong impact on the
following discussion as we are forced to combine data taken from
different cameras, and from the same instrument operated at different 
temperature conditions. 
Since the whole problem of absolute calibrations
(both with FOC and WFPC2) is under investigation, it is natural
to conclude that any differential variation of the zero-points
for one of the used configurations will affect the final results.

\section {The data}
\subsection {The Color Magnitude Diagrams}

The CMDs eventually obtained from all the measured stars in the
8 clusters are reported in Figure 4$a-h$. The clusters are shown in 
order of increasing metallicity (see Table 3, col. 4) 
to put into evidence the
regular trend in the variation of the CMD morphology with varying \fe.
In fact, as one can easily see from Figure 4, the slope of the
giant branch decreases progressively while the HB runs from the
blue for G 219 --the most metal-poor cluster-- to the red for 
G 280 --a cluster with almost solar \fe. 
\placefigure{fig4}

As already pointed out by \cite{ajh96} and \cite{ric95}, this 
overall behaviour makes the CMDs of the observed M31 GCs essentially identical 
to those of Galactic globulars having the same \fe. 
In particular, it is remarkable to note the increasing bending of
the bright giant branch with increasing metallicity, which leads
the most metal rich objects to display the well-known 
turn-over towards very red colors (\cite{ort90,bic91,ort92}). 

Since our specific aim here is to determine the \mv \vs \fe relation, 
we postpone any further quantitative analysis and discussion of the CMDs,
and simply report the procedure followed to derive the quantitites involved 
in this calibration.

\subsection {The cluster ridge lines}

To properly define meaningful ridge lines one must clean the cluster sample 
from contaminating objects and select a sub-sample of stars
measured in annular regions appropriate to optimize the stellar 
statistics with respect to the photometric quality (i.e. maximum number 
of stars with minimum blending and crowding problems). 

We have therefore plotted radial CMDs over various (4-6) annuli centered 
on each cluster and selected the best region to trace the ridge lines
of the various branches. Since the clusters present different levels
of background contamination (from G 219, where it is almost inexistent,
up to G 108, which is strongly contaminated) and different degrees 
of concentration, the definition of the ``best annulus'' varies from 
cluster to cluster. 

We list in Table 2 the ridge lines adopted for the various
clusters. We note that, in general, the ridge line traced
using the more internal annuli yields slightly bluer colors (by
$\sim 0.01$ mag) at fixed luminosity. This may be due to residual
field contamination in the more external areas, to blending effects
(along the RGB, brighter stars are redder than fainter ones, and the blend
is slightly brighter and bluer than the original bright component) or to 
a true color gradient across the cluster. Since we do not have any
suitable tool to disentangle these aspects, we have made our own best
choice, which may be uncertain to $\pm0.02$ mag at least.
Moreover, due to statistical reasons, it is quite
difficult to firmly trace the brightest part of the giant branch
as only few very bright stars actually populate the different annuli.

\placetable{tbl-2}

\subsection {Adopted reddenings and metallicities}

Estimates of reddening and metallicity for the program clusters are 
already available in the literature or can be derived from available data. 
The simplest and most straightforward procedure is then to adopt these 
values, which are independent of the \hst observations.

Table 1, col. 10 reports the estimates of the reddening --E(B-V)--
we made from the HI maps (\cite{bur82}, Figure 4g).
They are substantially identical to the values reported by \cite{ajh96} 
and \cite{ric95}, based on the same procedure . Though
alternative estimates can be derived (see below), we have adopted these 
values for E(B-V) in the calculation of the luminosity-metallicity 
calibration. 

Concerning metallicity we note that, while it has been impossible so 
far to get CMDs from ground-based observations of the M31 GCs, several 
estimates of their metallicities could be obtained from spectroscopy and
multi-color photometry of the cluster integrated light (see \cite{huc91,bon87}
for further references).

We report in Table 1 (col. 7, 9) the values of \fe for the program 
clusters obtained by \cite{huc91} from a set of spectroscopic indices, 
and those reported by \cite{bon87} from V-K integrated colors 
calibrated in terms of \fe. These two estimates are totally
independent and their major difference is that the former
are reddening independent, while the latter require the knowledge
of the individual reddenings.  Apart from the quality of the 
observational data, another source of error and/or systematic differences 
is also due to the different calibrations in
terms of \fe of the adopted indicators. In col. 8, we report
the value of the metallicity obtained from the same original V-K data
but using a revised calibration (Federici et al. 1996, in preparation). 
In general, all these estimates 
are in sufficiently good agreement for most clusters, although 
some more information seems to be essential in a few cases. 
In particular for G280 the photometric and spectroscopic metallicities 
are significantly different, while the revised calibration of the V-K data 
yields an intermediate value, which has then been adopted. 

On the other hand, the availability of new, sufficiently reliable ridge 
lines for the giant branches allows for the first time to estimate the 
metallicity also by comparing  the morphology and slope of the RGB of each 
individual M31 GC with a reference grid of Galactic GCs with known 
metallicities in the same observational plane (see for instance
\cite{dac90}, who applied this procedure 
to 6 globular clusters in the Galaxy). 
As already done by \cite{ajh96} and further
discussed in Sect. 3.4, the direct
comparison of the morphology of the RGB (made by searching for the
best match of the the ridge lines, independent of any constraints
on reddening and distance) can yield very useful indication on the
cluster metal content, for instance by considering the slope and
degree of bending of the giant branch. 

This comparison essentially confirms for all the clusters the reliability 
of the adopted values  and the ranking, except maybe for G1 (see Sect. 3.4).
However the {\it absolute} location of the RGB in the (M$_V$,B-V) plane 
(or equivalent) depends also on the reddening and distance of the cluster, 
and disentangling the effects of these parameters is not a trivial 
task, especially when the observational errors are still rather large. 
We have therefore decided to adopt for the metallicities the values
reported in Table 3, col. 4, and use the comparison with a reference
grid of Milky Way globulars just as a qualitative check (see Sect. 3.4).

\placetable{tbl-3}

Alternative procedures are also feasible, though less straightforward and 
accurate, which for instance may be useful for Bo468, never observed 
so far to get metallicity estimates.
In particular, one could apply the method developed by Sarajedini (1994,
hereafter {\it S-M}). This method would in principle allow to determine \fe and 
E(V-I) simultaneously from the giant branch. However, as noted by \cite{ajh96},
higher precision photometry than provided by the current data is 
really needed for this method. Nonetheless, the use of this procedure yields
estimates fully consistent within the errors with the adopted values 
for all considered clusters. In particular, for Bo468, the value of metallicity
reported in Table 3 (col. 4) has been obtained by adopting
E(B-V)=0.06 from the HI maps and then using the {\it S-M} to get \fe. 
Note that without any assumption on E(B-V), the {\it S-M} would converge
for Bo468 to E(B-V)$=0.07\pm0.04$ and \fe$=-0.61\pm0.40$.

\subsection {A reference grid: comparison with the Milky Way globulars}

As explained above, the comparison of the morphological shapes of the RGBs 
with a grid of reference clusters of known metallicities can be used as 
a further ``quality check'' of the adopted [Fe/H], or to get useful indication
in those cases where the spectroscopic and photometric determinations disagree 
significantly (e.g. G280) or are missing altogether. 

In this procedure the weakest points are the lack of precise measures
of individual reddening, that can be known only roughly with an error 
hardly smaller than about 0.03 mag, and the impossibility to account for some 
distance dispersion, which would be about $\pm$ 0.05 mag if the space 
distributions of the clusters is $\sim20$ Kpc.
Note that, assuming A$_V=3.2$E(B-V), the above uncertainties produce a
"vertical" error of about 0.10 mag in addition to the observational 
error of the HB itself (see Table 3 and Sect. 4). 

Adopting the values of reddening and metallicity reported in Table 3
(col.s 3, 4), and on the assumption that all M31 clusters are at the 
same distance (with adopted distance modulus to M31 \dmo=24.43, \cite{fre90}),
one can shift the RGB ridge lines in color and magnitude 
according to the adopted absorption law (E(V-I)=1.342(B-V), A$_I=$1.858E(B-V), 
see for reference \cite{ajh96}) to get the {\it absolute} location of 
each individual cluster ridge line with respect to the reference grid 
based on the Milky Way globulars.

We present in Figure 5 the results we have obtained in the (V,V-I)$_o$--plane 
for the 6 clusters observed with the WFPC2, and in Figure 6 
those obtained in the (B,B-V)$_o$--plane for the 2 clusters observed with 
the FOC. 
For summary and clarity, Table 4 reports  the values of the various 
parameters eventually adopted for the M31 clusters and those for
the Galactic reference clusters, whose data-base is briefly discussed
below.
\placetable{tbl-4}

To ensure homogeneity with \cite{dac90}, we have
adopted the relation \mv=0.17\fe+0.82 to set the distance-scale
zero-point and dependence on \fe of the "standard candle" (\mv).
In the (V,V-I)$_o$--plane the Galactic data are taken from \cite{dac90},
(Table X), to which we added the (V,V-I)$_o$ data of NGC 
6528 from Ortolani \etal (1992) and Guarnieri (1996, private comunication).
In the (B,B-V)$_o$--plane we reported the ridge lines of M92
(\cite{buo85,ste88}), M3 (\cite{buo94}), 47 Tuc (\cite{hes87}), 
and NGC 6553 (\cite{ort95}, Lanteri-Cravet, 1996, private
communication) appropriately 
scaled by adopting the distance scale from \cite{dac90},
and the metallicity scale from \cite{zin85} and \cite{arm89}.
\placefigure{fig5}
\placefigure{fig6}
As one can see from the Figures 5 and 6, the global matching is
sufficiently good in both planes. This implies that the basic
properties of the M31 clusters closely resemble those of the
Galactic globulars of similar metal content. Of course, the 
agreement with the reference grid in the \mv \vs color plane
depends quite strongly  also on the \mv \vs \fe relation assumed to 
derive the \mv values for the reference Galactic clusters. 
Furthermore, small residual discrepancies can be ascribed to
errors in the definition of the ridge lines and in the unknown
distance dispersion of the GCs within M31. One notable exception
seems to be the cluster G1, whose photometric and spectroscopic
indices indicate a somewhat lower metallicity than suggested
by its morphology in the I$_o$, (V-I)$_o$ plane (see Figure 5),
which is similar to 47 Tuc. On the basis of this comparison alone,
one could conclude that the metallicity of G1 is \fe $\sim -0.7$
(Rich \etal 1995). Since we decided to use this morphological
comparison only as a quality check of metallicity, we have
adopted the lower spectroscopic value of metallicity, \fe = -1.08
for consistency with the criteria we have applied on all the other
clusters. However, we must keep in mind this discrepancy, and we
shall see what impact it may have on our final luminosity--metallicity
calibration (see Sect. 4).

\subsection {The measure of the observed \vhb}

Concerning specifically the measure of \vhb, we have adopted various
approaches to take into account the intrinsic differences in the 
HB-morphologies, as done also by \cite{ajh96}. In
particular, for the metal-poor clusters with sufficiently extended HBs 
({\it i.e.} G219, G351, G105) we have computed a running mean over 
a 0.2 mag box moving along the HB in color, and we have adopted the value
at V-I$\sim0.5$, corresponding approximately to the center of the 
instability strip (see Table 3, col. 5). The associated uncertainty
is the observed {\it rms} vertical scatter of the HB at that color
(see Table 3, col. 9).

For the metal rich objects, where the HB collapses to a red clump of stars,
the identification of a reliable uncontaminated
HB sample is extremely difficult. Besides applying the technique of the
moving box adopted for the metal-poor clusters, we have also studied
the Luminosity Function (LF) to locate the HB clump. The use of the LF
is indeed made difficult by the fact
that for metal-rich clusters the HB is also partially merging with the 
so-called RGB-bump (\cite{tho67}), and the linear fit in Log of the
RGB LF is actually broken in correspondence of the RGB-bump 
(\cite{fp190}). Due to this fact, the procedure of
removing RGB stars from the LF as done by \cite{ajh96} may
be uncertain, though indicative. By coupling the two procedures
we have eventually drawn the estimates reported in Table 3 (col. 5).
The associated errors represent the observed {\it rms} scatter in the
adopted box (see Table 3, col. 9). 

A final general remark may be useful concerning the uncertainties
affecting the adopted \vhb values. In fact, looking at the CMD's
presented in Figure 4, one gets the impression that the HB is
clearly visible for all the clusters in the top panel, while it
is difficult to identify for instance for Bo 468 and G 280.
Consequently, one might expect to find much larger errors in the
measure of \vhb in the two latter cases. Actually, as shown in Table 8,
column 9, the uncertainties we have determined for these clusters
are slightly larger (0.04-0.05 mag) but not so much. This is
essentially due to the ability of the procedures we adopted to
locate fairly well the peak in the distributions in luminosity even if the
color range of the HB is very narrow, giving the impression of a uniform 
clump of stars.

\section {The \mv \vs \fe calibration}

Having determined the observed \vhb we can now consider the calibration
of the standard candle. We adopt \dmo = 24.43 (\cite{fre90}) but other 
distance moduli spanning a range of at least
$\pm0.2$ mag can be found in the literature (see for instance  
\cite {pvdb88,chr91}).
While this choice has no effect on the following discussion concerning
the slope of the relation, it is crucial for the comparison of
the zero-points and, in turn, for the problem of {\it absolute}
ages. Note that an uncertainty of $\sim 0.07$mag in the zero-point
implies a corresponding uncertainty of $\sim 1$ Gyr in age.

Most of the impact on the determination of the slope comes from two
basic items, {\it i.e.} the need for a correction of the observed 
\vhb to take into account the influence of the HB-morphology variations
with varying metallicity, and the treatment of the
{\it global} errors associated to the individual \mv and \fe values 
adopted for each considered cluster. Note that, as explained below,
the first item affects mostly the slope, while the second one
essentially drives its associated error.

Concerning the first aspect, it is well known that
metal-rich GCs in the Milky Way do not 
contain any HB star a part from those populating a red stubby clump,
which is however slightly brighter (by a quantity \drich) than the average 
luminosity RR Lyrae variables would have, if they existed in these metal-rich 
clusters, according to stellar evolution theory
(see for instance  \cite{lee94}) .
While the need for such a correction is generally accepted, its absolute 
size is a matter of discussion and it largely depends on theoretical
models.
Following \cite{sar95}, \cite{ajh96} and \cite{cdf96}, 
we have adopted a correction \drich$=0.08$mag to
the values of the observed \vhb (Table 3, col. 5) for the clusters 
with \fe $> -1.0$  (see col. 7). This represents a compromise among 
possible values ranging from $\sim 0.05$ to 0.12 mag, as reported 
in the above quoted studies. 
According to the theoretical ZAHB models by \cite{swe87}, \cite{ccp91} 
and \cite{dor92}, 
a value around 0.09-0.10 would be appropriate (see also Fullton et al. (1995), 
however we have adopted 0.08 for consistency with \cite{ajh96}.

In Sect. 3 we have discussed several possible error sources, 
which include contributions from:  a) the adopted 
measure of \vhb (due to photometric errors, crowding, statistics, etc.); 
b) aperture photometry corrections and differences between the two cameras; 
c) the photometric transformations and the definition of the 
zero-points; d) the individual reddenings; e) the adopted reddening law; 
f)  the metallicity estimates, etc. 
Moreover, we have assumed that the clusters are located 
at the same distance to us. The reliability of this 
assumption is actually the "intrinsic" strength of this procedure. 
However, this is not true in reality, and (small, i.e. $\le \pm$ 0.05 mag) 
differential distances also contribute to the scatter in the derived relation.

It is therefore quite hard to quantify precisely a fair measure of the 
{\it global} error one should associate to each individual value while
combining all the available data to yield a final \mv \vs \fe
calibration.

To reach a conclusion (at least provisional, until new information
will become available) we have decided to add in quadrature to
each formal error obtained from the procedures described at
Sect. 3.5 (see Table 3, col. 9) an additional error of 0.10 mag which 
should statistically include the effects of all the above mentioned 
sources of uncertainty, allowing also for the fact that some of them 
could compensate each other to some extent. The final figures so 
obtained and adopted for
the {\it global} error (as 1$\sigma$ error-bar) are listed in Table 3, 
col. 10.

In Figure 7 we show the best linear fit obtained by taking into 
account the size of the errors on both variables simultaneously, and the 
two ($\pm1\sigma$ ) solutions of the slope coefficient, 
normalized at \fe=$-$1.1. 
The final \mv \vs \fe  relation so obtained is:
\begin{displaymath}
M_{V}^{HB} = (0.13 \pm 0.07)\fe + (0.95 \pm 0.09)
\end{displaymath}

\placefigure{fig7}

This result is of course strongly dependent on the various assumptions
(more or less explicitly) made during the whole procedure, and should
be used with care. The following considerations may thus be useful:
\begin{enumerate}
\item The {\it zero-point} is totally dependent on the adopted distance
modulus to M31 and, moreover, it depends on the zero-points of the
\hst photometric calibrations (which are uncertain to $\sim\pm0.05$mag).
As already pointed out, this aspect is crucial to any use of this
relation to get distances and {\it absolute} ages of globular clusters,
and we recall that a variation of $\sim +0.07$  mag implies an age
increase of $\sim +1$Gyr.
\item The {\it slope} and its accuracy depend on several different 
aspects, which can be schematically summarized as: 
(a) systematic differential
errors in the zero-points of the various camera configurations;
(b) systematic differential errors in the measurement of \vhb
with varying HB morphology, like for instance the correction for the red 
HBs. If this correction were assumed to be $\sim$ 0.10 mag the slope 
of this relation would become 0.15$\pm$0.07; 
(c) errors in the metallicity estimates. In this respect, assuming for 
instance for G1 \fe$=-0.70$ rather than -1.08
(see Sect. 3.4), the slope of the relation would be 0.12 $\pm$ 0.07; 
(d) distance dispersion within M31.
If one wanted to be extremely conservative, one could adopt an error of 
$\pm$0.15 mag for \vhb for all clusters, and one would then obtain: 
\begin{displaymath}
M_{V}^{HB} = (0.13 \pm 0.10)\fe + (0.95 \pm 0.12)
\end{displaymath}
\end{enumerate}

To have further insights on the intrinsic reliability of the \mv \vs \fe
relation here obtained, it could also be useful to examine 
homogeneous  subsamples of clusters obtained with the same camera setup.
In particular, one would get the following relations:
\begin{displaymath}
M_{V}^{HB} = (0.07 \pm 0.10)\fe + (0.86 \pm 0.11)
\end{displaymath}
from the 6 GCs observed with the WFPC2, irrespective of the camera operating 
temperature;
\begin{displaymath}
M_{V}^{HB} = (0.07 \pm 0.10)\fe + (0.87 \pm 0.12)
\end{displaymath}
from the 5 GCs observed with the WFPC2 at constant temperature ($-88^o$C);
and
\begin{displaymath}
M_{V}^{HB} = (0.21 \pm 0.11)\fe + (1.11 \pm 0.14)
\end{displaymath}
from the 2 GCs observed with the FOC.
It is difficult to give more weight to the individual sub-sets than to
the whole sample as the number of observed clusters is too low in all cases. 
However, considering the various results,
it seems quite reasonable to conclude that the slope of the
\mv \vs \fe relation one can deduce at this stage from the analysis
of the M31 GCs is probably closer to 0.15, the valued predicted
by theoretical models (\cite{swe87,lee94}) 
and by several studies using ground-based observations 
(see for references \cite{lee94}), 
than to $\sim0.30$ (or even larger) that was obtained from other methods
(Sandage 1982, 1993). In this respect, a final comment 
may be worth of note. The use of the de-reddened \vhb values without adopting
any correction for the HB-morphology dependence ({\it i.e.} assuming
\drich$=0.00$) would yield over the whole sample:
\begin{displaymath}
M_{V}^{HB} = (0.06 \pm 0.07)\fe + (0.84 \pm 0.09)
\end{displaymath}
Although there is no doubt that the HB-morphology correction is needed, 
this confirms that the available data hardly support values  
as high as 0.30 or more for the slope of the \mv \vs \fe relation. 

\section {Conclusions and future prospects}

The study of the \mv \vs \fe relation carried out using the first
available data on the observations of M31 GCs with both imaging
cameras of \hst leads to conclude that there is no evidence
of any significant discrepancy between
the observational result and the HB-model predictions.

A wider sample of observed GCs in M31 is needed in order to reduce 
significantly the errors and confirm unambiguously the value of the slope 
derived in the present study, i.e. $\sim$ 0.13. 

In particular, it would be very important to have data for clusters
displaying a well populated HB, over a wide color range and covering
an appropriate range in metallicity. A proposal by our team 
(GO 6671, P.I. M. Rich) has been granted \hst time to observe 10 more 
GCs in M31 during Cycle 6. 
As soon as these new data will be available, we are confident that
the long-standing issue of measuring the slope of the \mv \vs \fe
relation will be settled.

\acknowledgements
It is a pleasure to thank our collaborators P.~Battistini, F.~Bonoli,
I.R.~King, and R.~Walterbos for their contributions to this project.
This research was supported by the {\it Consiglio delle Ricerche Astronomiche}
(CRA) of the {\it Ministero delle Universit\`a e della Ricerca Scientifica 
e Tecnologica} (MURST), Italy, by the {\it Agenzia Spaziale Italiana} (ASI),
and by the National Aeronautics and Space Administration, and
the National Science Foundation, USA.

\newpage
\figcaption[]{FOC/96 image of the globular cluster G351 in M31. Filter F480LP,
 co-added frames, total exposure time 3800 sec. \label{fig1} }

\figcaption[]{Adopted empirical transformations from the HST magnitude system
to the Johnson standard system for the FOC/96. \label{fig2} }

\figcaption[]{Comparison of the CMD's of the globular cluster G105 in M31, 
based on the same WFPC2 original data. (a) this paper, (b) Ajhar et al. 1996.
\label{fig3} }

\figcaption[]{CMD's obtained from the HST observations for the considered
sample of 8 globular clusters in M31. The diagrams are ordered with
increasing metallicity [Fe/H] (see labels). \label{fig4}}

\figcaption[]{Comparison of the observed giant branches of a sample of
reference Galactic globular clusters (dashed lines) and the observed 
M31 globulars (full line) in the I$_{0}$,(V-I)$_{0}$ plane. The Galactic
globulars are M15, [Fe/H]=-2.17, NGC 6397, [Fe/H]=-1.91, M2, [Fe/H]=-1.58,
NGC 6752, [Fe/H]=-1.54, NGC 1851, [Fe/H]=-1.29, 47Tuc, [Fe/H]=-0.71,
NGC 6528, [Fe/H]=-0.14, from blue to red. \label{fig5} }

\figcaption[]{Comparison of the observed giant branches of a sample of
reference Galactic globular clusters (dashed lines) and the observed 
M31 globulars (full line) in the V$_{0}$,(B-V)$_{0}$ plane. The Galactic
globulars are M92, [Fe/H]=-2.24, M3, [Fe/H]=-1.65,
47Tuc, [Fe/H]=-0.71, NGC 6553, [Fe/H]=-0.20, from blue to red. \label{fig6} }

\figcaption[]{The adopted $M_{V}(HB)$ vs [Fe/H] relation obtained from the 
HST observations of the considered sample of 8 GC's in M31. The associated
uncertainties are $\pm$ 0.2 dex in [Fe/H] and the {\it global} limiting errors 
reported in Table 3, col. 10 (see Sect. 3.5 and 4).
The full line represents the weighted best fit M$_{V}$(HB) = 0.13 [Fe/H] + 0.95;
the dashed lines (normalized at [Fe/H]=-1.1) indicate the uncertainty in the
slope (0.13 $\pm$0.07). \label{fig7} }

\clearpage

\begin{deluxetable}{rrrccrccccc}
\footnotesize
\tablecolumns{10}
\tablewidth{0pt}
\tablecaption{Relevant data for the 8 globular clusters observed in M31 
 with HST. \label{tbl-1}} 
\tablehead{
\colhead{Bo\tablenotemark{a}} &\colhead{G\tablenotemark{a}}&
\colhead{Vet\tablenotemark{a}} &\colhead{V} &\colhead{(B-V)} &
\colhead{R$_{kpc}$\tablenotemark{b}} &\colhead{[Fe/H]$_p$\tablenotemark{c}} &
\colhead{[Fe/H]$_{pr}$\tablenotemark{d}} &
\colhead{[Fe/H]$_{s}$\tablenotemark{e}} &
\colhead{E(B-V)$_{BH}$\tablenotemark{f}} &
\colhead{E(B-V)$_{S}$\tablenotemark{g}} }
\startdata
   &   1& MII\phn& 13.70& 0.86& 34.16& -1.33& -1.14& -1.08& 0.05& 0.03 \nl
  6&  58&  55\phn& 15.60& 0.89&  6.32& -0.51& -0.72& -0.57& 0.10& 0.11 \nl
343& 105& 199\phn& 16.35& 0.69& 14.52& -1.25& -1.11& -1.49& 0.06& 0.04 \nl
 45& 108&  94\phn& 15.77& 0.88&  4.81& -0.86& -1.08& -0.94& 0.12& \nodata \nl
358& 219& MIV\phn& 15.11& 0.60& 19.53& -2.28& -1.90& -1.83& 0.06& 0.03 \nl
225& 280& 282\phn& 14.26& 0.92&  4.59& -0.21& -0.44& -0.70& 0.09& \nodata \nl
405& 351& 140\phn& 15.16& 0.72& 17.71& -1.95& -1.82& -1.80& 0.10& \nodata \nl
468&    &    & 18.12& 0.53& 19.80&\nodata &\nodata & \nodata & 0.06& 0.07 \nl
\enddata
\tablenotetext{a}{Bo from Battistini et al. (1987), G from Sargent et
al. (1977), Vet from Vetesnik (1962)}
\tablenotetext{b}{adopting for M31 a distance of 770 kpc, corresponding to
$(m-M)_0$ = 24.43}
\tablenotetext{c}{metallicities from infrared colors $(V-K)_{0}$ (B\`onoli et 
al. 1987)}
\tablenotetext{d}{metallicities from $(V-K)_{0}$, using $E(B-V)_{BH}$ and the
calibration from Brodie \& Huchra (1990)(Federici \phn\phn\phn et al. 1996, 
in preparation)}
\tablenotetext{e}{metallicities from spectral indices (Huchra et al. 1991)}
\tablenotetext{f}{reddenings from the HI maps (Burstein \& Heiles 1982)}
\tablenotetext{g}{reddenings obtained applying the method developed by 
Sarajedini (1994)}
\end{deluxetable}

\clearpage

\begin{deluxetable}{ccccccccccc}
\footnotesize
\tablecolumns{11}
\tablewidth{0pc}
\tablecaption{Ridge lines for the 8 globular clusters observed in M31 
 with HST. \label{tbl-2}} 
\tablehead{
\multicolumn{2}{c}{G219}&\colhead{}  &\multicolumn{2}{c}{G351}&\colhead{} &
\multicolumn{2}{c}{G105}&\colhead{}  &\multicolumn{2}{c}{G1} \nl
\cline{1-11}\nl
\colhead{I} &\colhead{(V-I)} &\colhead{} &
\colhead{V} &\colhead{(B-V)} &\colhead{} &
\colhead{I} &\colhead{(V-I)} &\colhead{} &
\colhead{I} &\colhead{(V-I)} }
\startdata
20.50& 1.52& & 22.34& 1.70& & 20.50& 1.73& & 20.75& 2.40\nl
21.00& 1.37& & 22.50& 1.58& & 21.00& 1.53& & 21.00& 2.00\nl
21.50& 1.28& & 23.00& 1.36& & 21.50& 1.40& & 21.50& 1.64\nl
22.00& 1.20& & 23.50& 1.22& & 22.00& 1.29& & 22.00& 1.45\nl
22.40& 1.16& & 24.00& 1.10& & 22.50& 1.21& & 23.00& 1.20\nl
23.00& 1.09& & 24.50& 1.00& & 23.00& 1.15& & 23.50& 1.12\nl
23.45& 1.05& & 25.00& 0.92& & 23.50& 1.10& & 24.00& 1.07\nl 
24.00& 1.01& & 25.50& 0.89& & 24.00& 1.05& & 24.50& 1.02\nl 
25.00& 0.94& & 26.00& 0.88& & 25.00& 1.00& & 25.00& 0.95\nl
\nl
\cline{1-11} \nl
\multicolumn{2}{c}{G108}&\colhead{}  &\multicolumn{2}{c}{Bo468}&\colhead{} &
\multicolumn{2}{c}{G58}&\colhead{}  &\multicolumn{2}{c}{G280} \nl
\cline{1-11}\nl
\colhead{I} &\colhead{(V-I)} &\colhead{} &
\colhead{I} &\colhead{(V-I)} &\colhead{} &
\colhead{I} &\colhead{(V-I)} &\colhead{} &
\colhead{V} &\colhead{(B-V)} \nl
\cline{1-11}\nl
20.75& 2.20& & 21.00& 2.00& & 21.00& 3.00& & 25.00& 2.40\\
21.00& 1.90& & 21.30& 1.80& & 21.50& 2.00& & 24.50& 2.17\\
21.50& 1.68& & 21.75& 1.60& & 21.65& 1.80& & 24.00& 1.90\\
22.00& 1.54& & 22.45& 1.40& & 22.00& 1.65& & 23.55& 1.68\\
22.50& 1.43& & 23.60& 1.20& & 22.50& 1.50& & 23.50& 1.64\\
23.00& 1.35& & 24.00& 1.15& & 23.00& 1.35& & 23.55& 1.60\\
23.30& 1.30& & 24.40& 1.10& & 23.50& 1.27& & 24.00& 1.45\\
24.00& 1.20& & 24.95& 1.05& & 24.00& 1.20& & 24.50& 1.33\\
24.45& 1.17& & 25.30& 1.00& & 24.52& 1.17& & 25.00& 1.26\\
25.00& 1.13& & 25.80& 0.95& & 25.00& 1.15& & 25.50& 1.20\\
\nodata &\nodata& & \nodata&\nodata& & \nodata&\nodata& & 26.00& 1.10\\
\enddata
\end{deluxetable}

\clearpage

\begin{deluxetable}{rrcccccccccc}
\footnotesize
\tablecolumns{12}
\tablewidth{0pc}
\tablecaption{Adopted reddenings, metallicities and Horizontal Branch 
magnitudes \label{tbl-3} }
\tablehead{
\colhead{Bo} &\colhead{G} &\colhead{} &\colhead{E(B-V)$_{ad}$}&
\colhead{[Fe/H]$_{ad}$} & \colhead{V(HB)} &
\colhead{V$_{0}$(HB)} &\colhead{V$_{0}$(HB)$^{cor}$} &
\colhead{M$_{V}$(HB)$_{0}^{cor}$} &\multicolumn{3}{c}{$\varepsilon$V(HB)} \nl
\colhead{} &\colhead{} &\colhead{} &\colhead{} &\colhead{} &\colhead{} &
\colhead{} &\colhead{} &\colhead{} &\colhead{*} &\colhead{} &\colhead{+} }
\startdata
   &   1& & 0.05\phn& -1.08\phn& 25.31& 25.15& 25.15& 0.72& 0.05& & 0.11 \nl
  6\phn&  58& & 0.10\phn& -0.57\phn& 25.45& 25.13& 25.21& 0.78& 0.03& & 0.10 \nl
343\phn& 105& & 0.06\phn& -1.49\phn& 25.49& 25.30& 25.30& 0.87& 0.03& & 0.10 \nl
 45\phn& 108& & 0.12\phn& -0.94\phn& 25.53& 25.15& 25.23& 0.80& 0.03& & 0.10 \nl
358\phn& 219& & 0.06\phn& -1.83\phn& 25.29& 25.10& 25.10& 0.67& 0.03& & 0.10 \nl
225\phn& 280& & 0.09\phn& -0.40\phn& 25.66& 25.37& 25.45& 1.02& 0.04& & 0.11 \nl
405\phn& 351& & 0.10\phn& -1.80\phn& 25.48& 25.16& 25.16& 0.73& 0.06& & 0.12 \nl
468\phn&    & & 0.06\phn& -0.61\phn& 25.40& 25.21& 25.29& 0.86& 0.05& & 0.11 \nl
\enddata
\tablenotetext{*}{formal rms--scatter in the adopted HB box (see Sect. 3.5)}
\tablenotetext{+}{global error (see Sect. 4)}
\end{deluxetable}

\clearpage 

\begin{deluxetable}{lcccc}
\footnotesize
\tablecolumns{5}
\tablewidth{0pc}
\tablecaption{Adopted values for the MW. \label{tbl-4}} 
\tablehead{
\colhead{Names} &\colhead{}& \colhead{$[Fe/H]_{ad}$} &\colhead{$E(B-V)_{ad}$} &
\colhead{$(m-M)_{0}$} }
\startdata
M92 & & -2.24\phs& 0.02\phs & 14.46\phs \nl
M 15 & & -2.17\phs& 0.10\phs & 15.41\phs \nl
NGC 6397 & & -1.91\phs& 0.18\phs & 12.40\phs \nl
M3 & & -1.65\phs & 0.00\phs & 14.94\phs \nl
M 2 & & -1.58\phs& 0.02\phs & 15.50\phs \nl
NGC 6752 & & -1.54\phs& 0.04\phs & 13.19\phs \nl
NGC 1851 & & -1.29\phs &0.02\phs & 15.45\phs \nl
47Tuc & & -0.71\phs & 0.04\phs & 13.51\phs \nl
NGC 6528 & & -0.14\phs & 0.60\phs & 14.43\phs \nl
NGC 6553 & & -0.20\phs & 0.80\phs & 13.65\phs \nl
\enddata
\end{deluxetable}

\clearpage

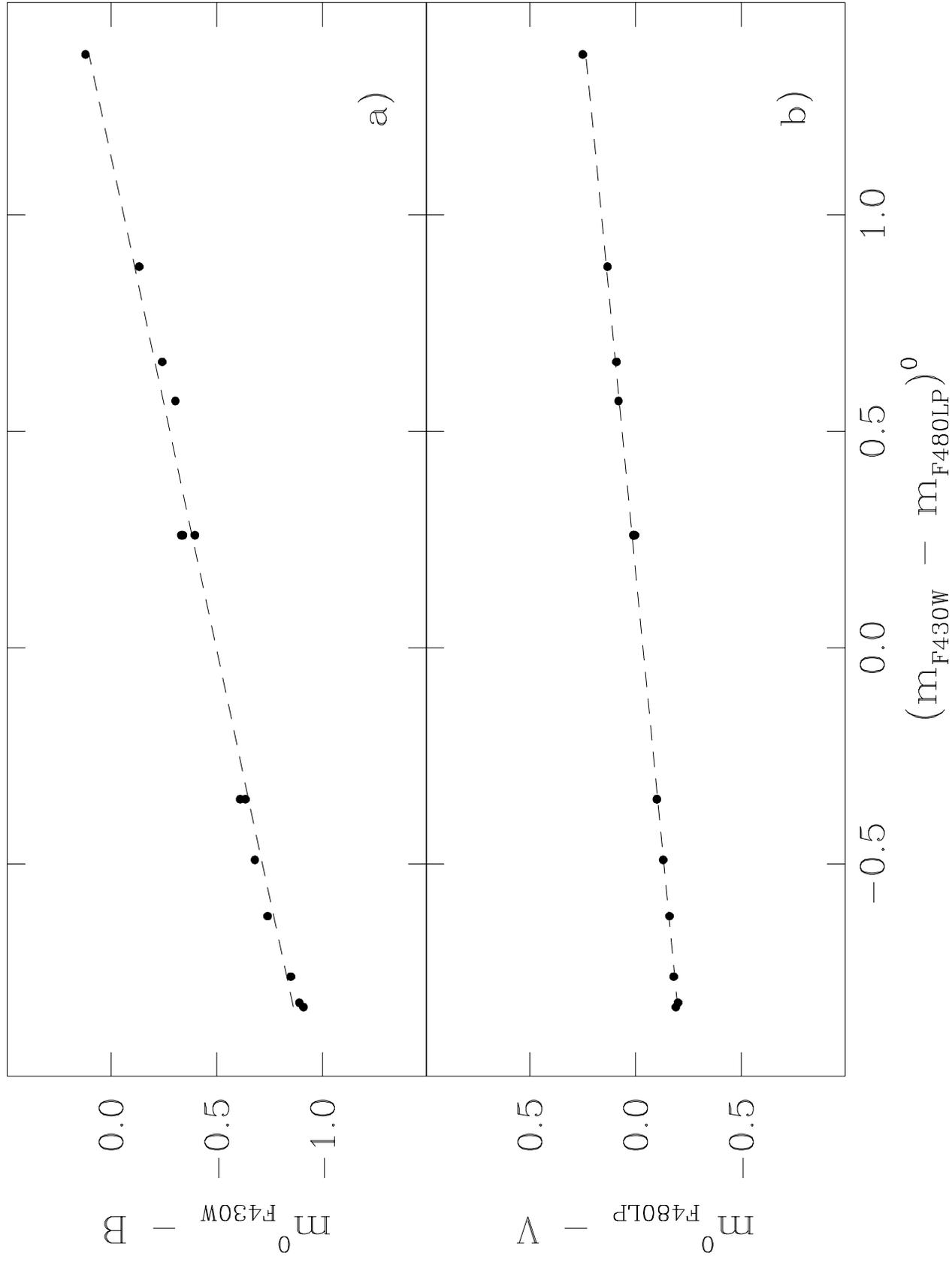
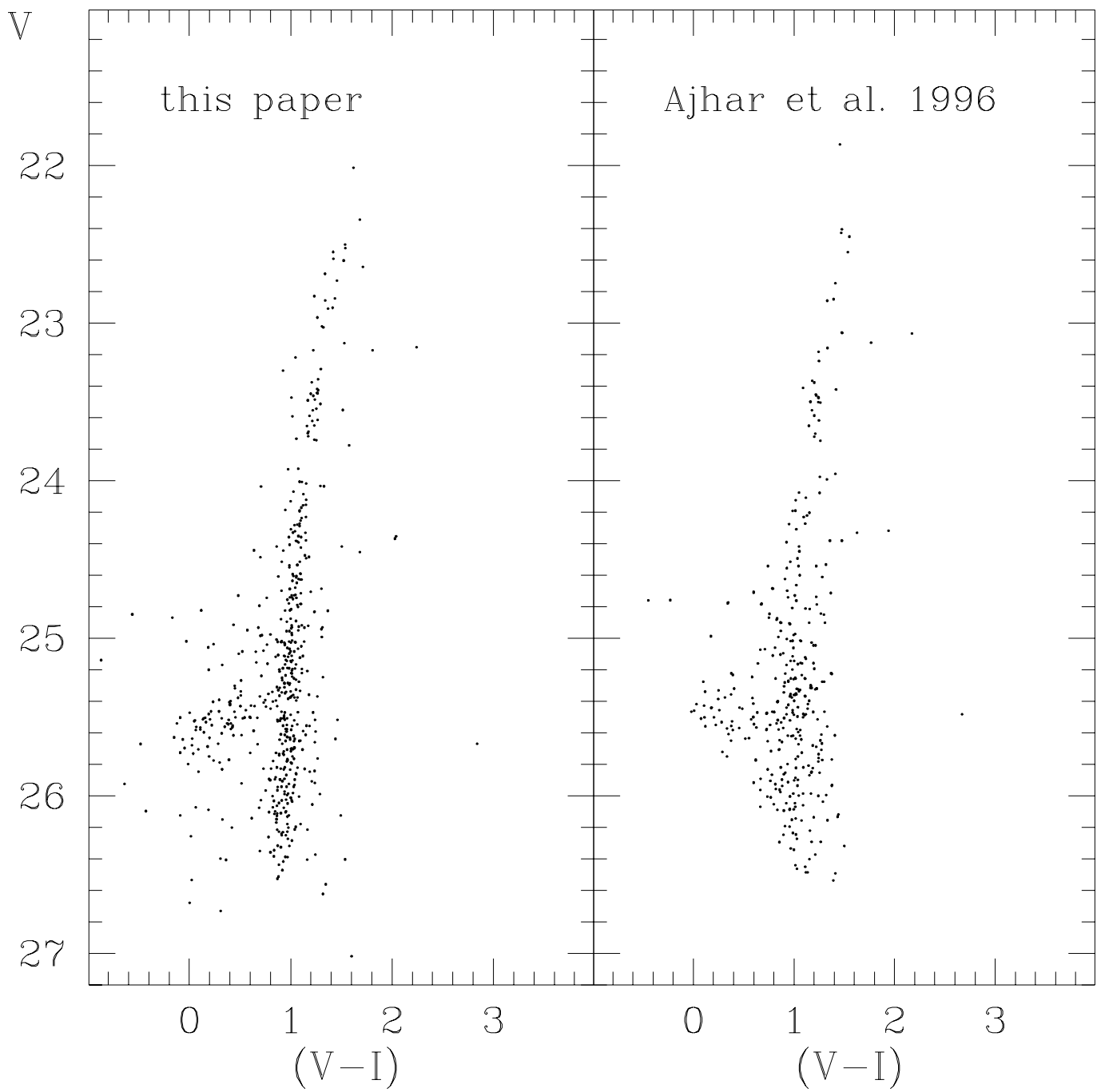
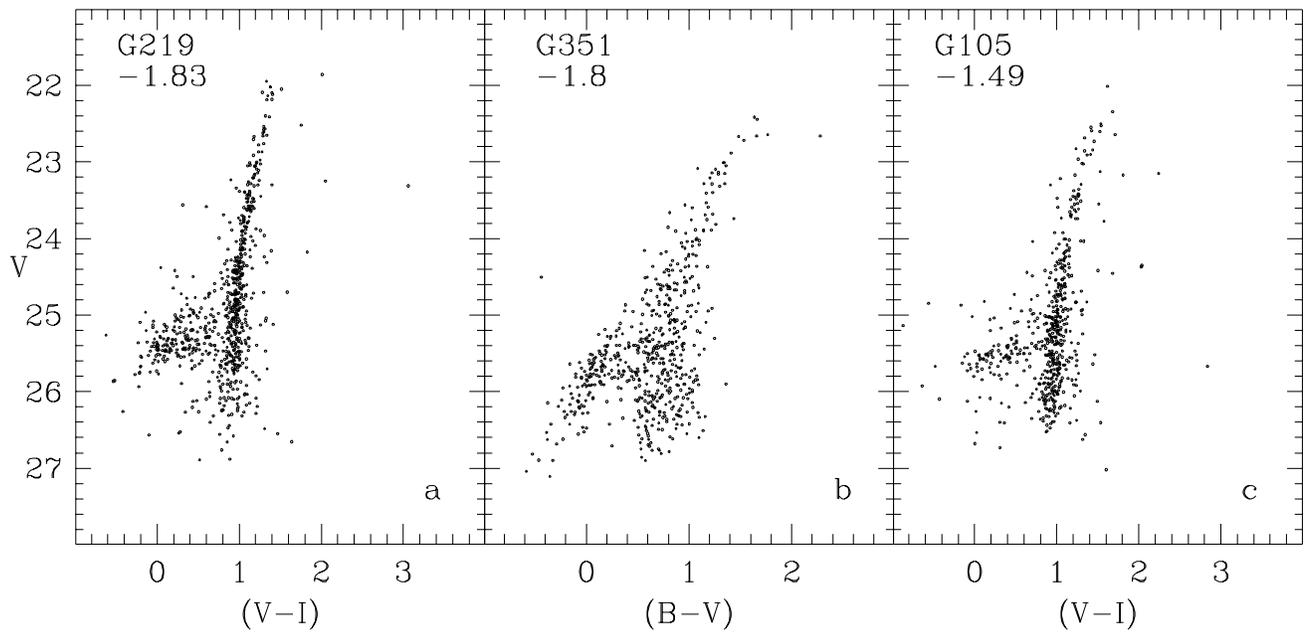
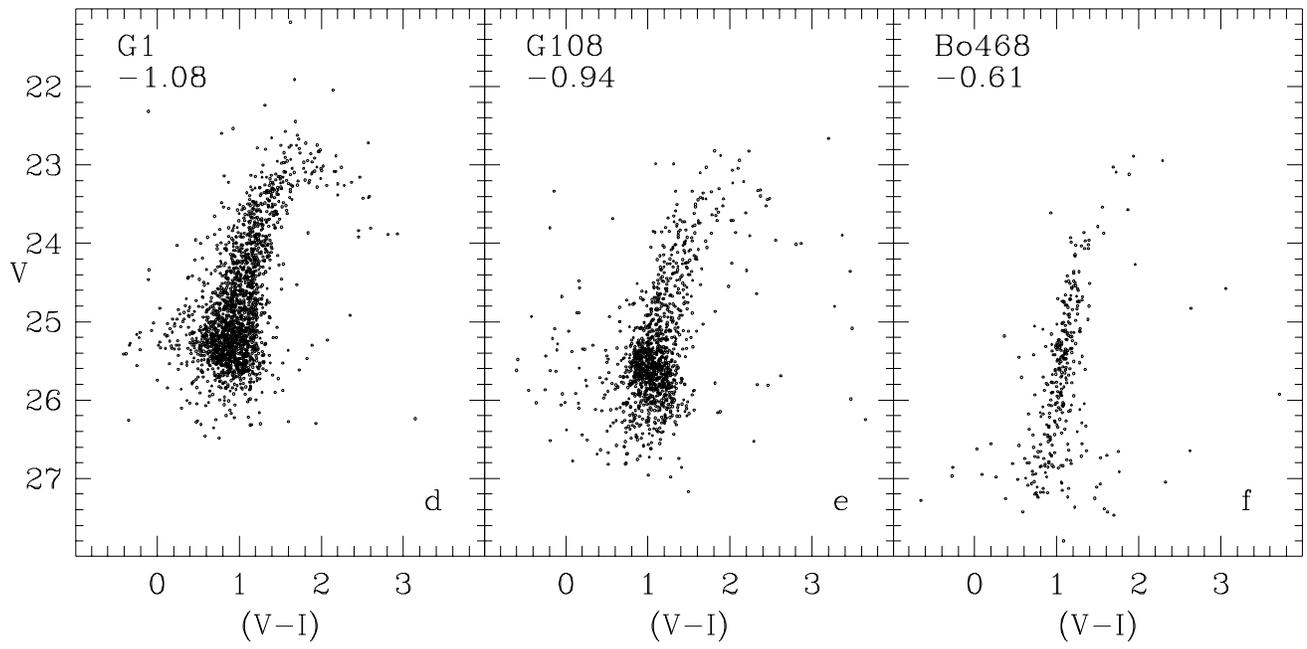
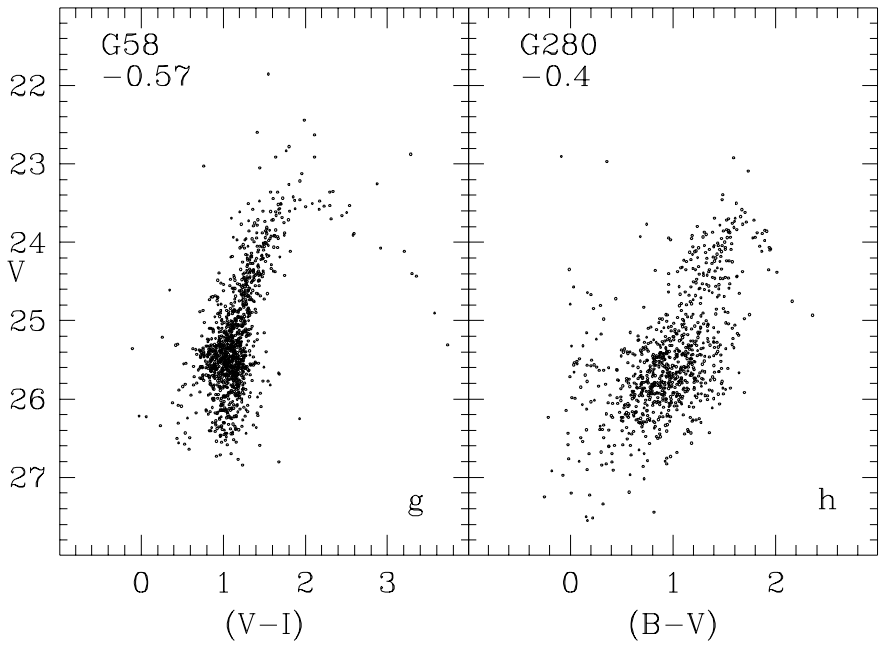
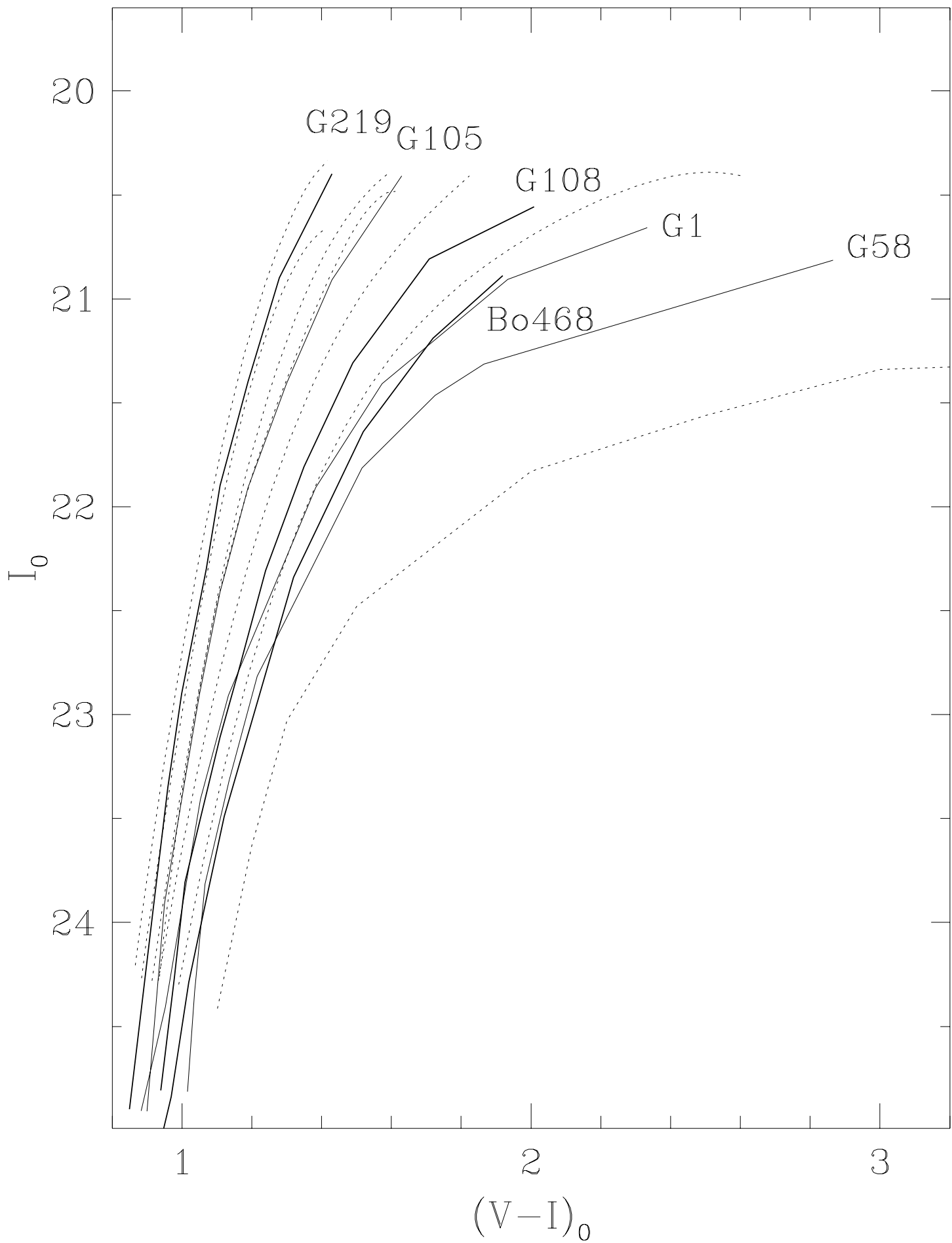
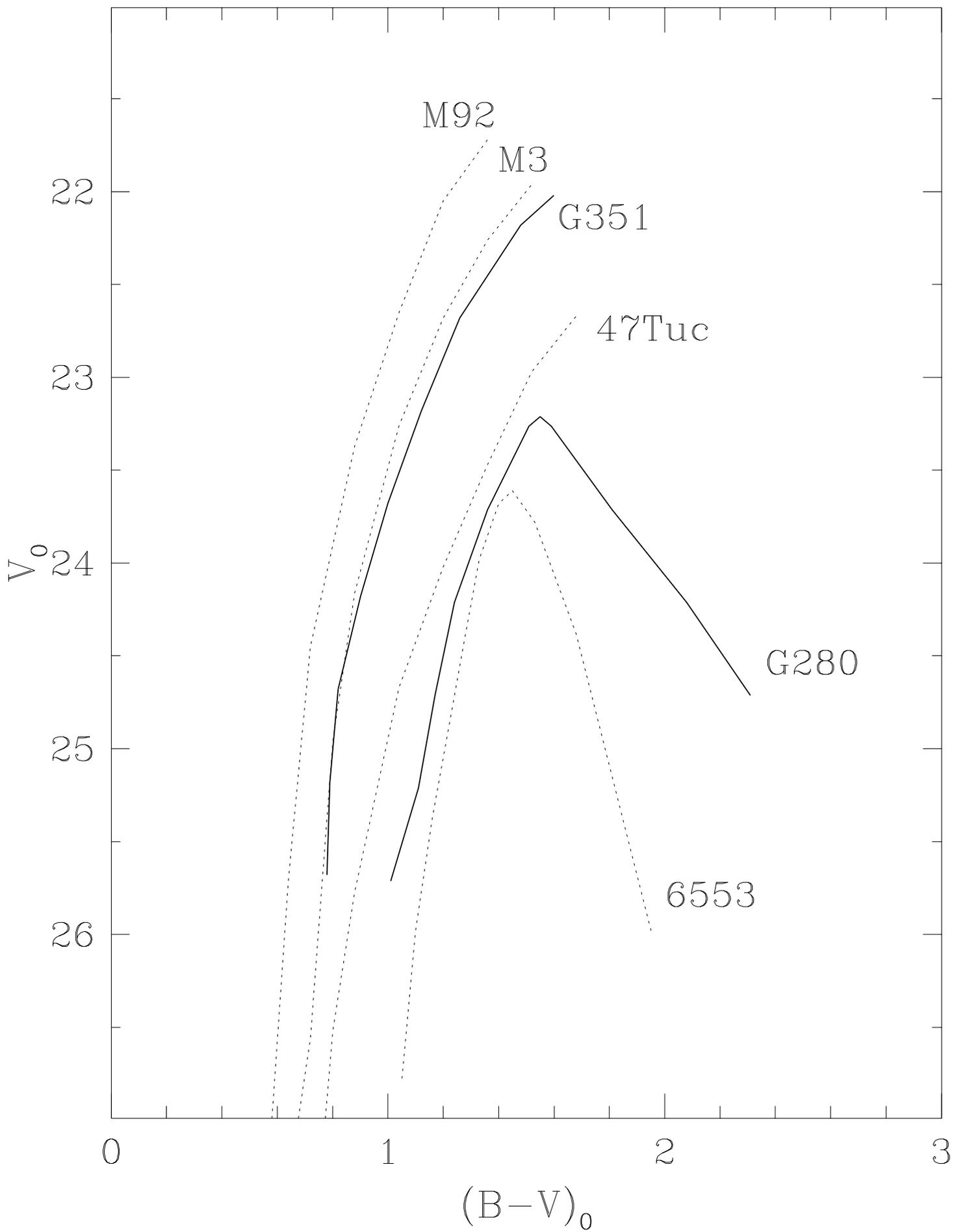
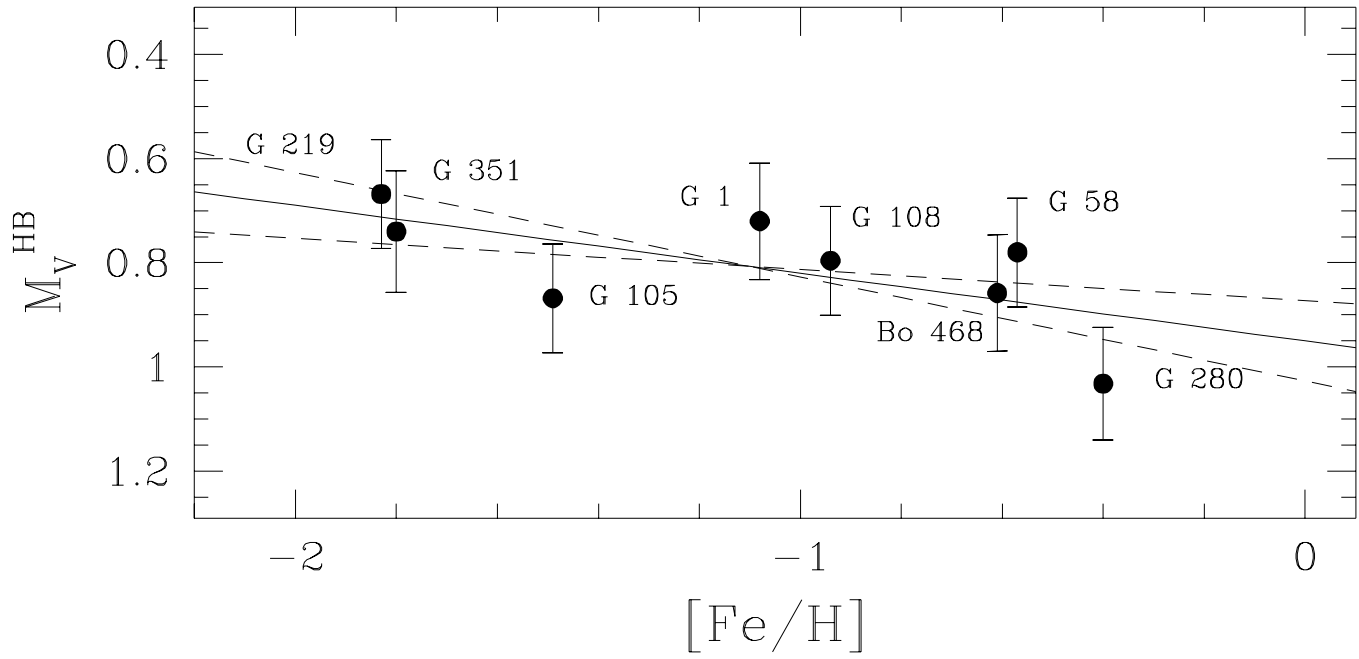
\end{document}